# Toward a new language of legal drafting

Matthew Roach
LLM 2015 Stanford University
May 29, 2015*

**Abstract**

Lawyers should write in document markup language just like web developers, digital publishers, scientists, and almost everyone else.

*Keywords:* legal drafting, contracts, document markup language, XML, HTML, LaTeX

## Contents

*Thanks to Yegor Tkachenko, MS 2016, Loek Janssen, MS 2016 and Kevin O'Connell, JD 2017 for their valuable comments and discussion.

# List of Figures

# List of Tables

## 1. The problem

### *1.1. What got me thinking*

A common experience of being a lawyer that you don't think much about process improvement or product design. The key focus for many lawyers is meeting client needs as quickly as possible and billable hour targets. Having been a transactional lawyer for several years, I had never thought of drafting contracts in anything other than Microsoft Word.

When I started my LLM I met math and engineering students, who were involved in various forms of data analytics, machine learning and natural language processing. They showed me their projects and the software tools they were using. I realized that in other disciplines, people are adept at switching between the languages of math, coding or natural language, often within a single document, in order to use the tool best adapted to the task at hand.

Taking classes in design, technology and law, I began to think about the potential for changing how we generate and access legal content. I began to reflect on how we access content in various forms through technology, and how far the design and accessibility of legal content lags behind what we now take for granted everywhere else.

This paper explores the thought that there is an enormous potential functionality that can be added to legal content if lawyers make modest efforts to add machine readable structure to their drafting. Lawyers would enjoy learning new skills, and clients and lawyers alike would be excited to discover how they way they produce and access legal content could be transformed.

This paper discusses what authoring in a markup language might look like, some of the advantages that this could have, and some of the barriers to implementation. A related question is what it would take to shift lawyer behavior to this style of writing, and what transitional steps might be appropriate. This could be the subject of further work.

### *1.2. How lawyers draft and publish contracts*

Lawyers draft documents in word processors that focus on formatting and final appearance, usually Microsoft Word. Their documents are almost universally accessible and editable by the lawyer's clients, the other side and the courts. Following initial preparation by a lawyer, a draft contract may be emailed back and forth many times, with the parties making and tracking various changes.

Once the parties agree the terms, a junior associate tidies up the formatting of the document, prints it out and walks around town getting it signed. If you're super modern, you might do electronic signatures. Then the associate makes a pdf, emails it around and everyone uses that or the final word document for ever after as the record of the deal struck.

### *1.3. Why this is a problem*

The output of legal drafting as it is done now and has been done in the past is unstructured natural language, poorly adapted for computational use and analysis. This is a problem because:

- Outside law, clients and lawyers are used to accessing and editing content in much more user friendly formats.[1] They use web based platforms that look great, allow easy navigation and transformation of content presentation according to the users needs. They can do this because the content has embedded structures readable by computers.

- Sharing and editing content using traditional word processors is cumbersome compared to tools used to create, edit and share digital content outside law. GitHub and similar tools used by coders offer much greater flexibility and functionality. There are significant productivity and user experience costs to keeping outdated tools just because they have become familiar and something of an industry standard in law.

- Companies don't know enough about the contracts they have entered into. As Nick West of Axiom has noted "very few general counsel can tell you the number of contracts their company is party to, let alone understand the totality of their obligations, the interactions between them or their organisational risk implications."[2]

- Lawyers aren't managing their knowledge and experience effectively. A significant part of the practice of law is drawing on experience and knowledge gained from previous transactions and documents.[3] Legal drafting often involves a lawyer taking a moderately well structured precedent[4] and customizing it into an unstructured or "flattened" form. It is hard to force the output back into the knowledge management system. In most cases the lawyer doesn't bother, which affects both the quality and efficiency of future work.

- While machine learning and natural language processing techniques are improving, the ability to undertake computational analysis of legal documents is significantly complicated by their lack of structure. If the lawyer is conscientious in how they use their firm styles it may be possible to parse a basic structure out of legal document and identify what are section headings, defined terms, legislation, case names, etc. But this is hard work and unreliable. It would be much better to start with something structured.[5]

---

[1] Although some of the new market entrants to the legal industry are showing how design and legal content can meet, see for example Rocket Lawyer's website at `www.rocketlawyer.com`, last visited May 28 2015.

[2] Nick West, *Getting to grips with corporate contracts*, The Lawyer, May 26 2015, `http://tinyurl.com/thelawyercorporatecontracts`.

[3] Stephen Choi et al. suggest that this is such a dominant aspect of drafting that contract drafters are more inventors than authors, taking existing products and try and improving them so that they can meet the clients' needs at hand, in Stephen Choi, Mitu Gulati and Eric Posner, *The dynamics of contract evolution* 88 New York University Law Review 1 (2013).

[4] For example, in my own law firm, the precedents collection has a machine readable structure which allows modular clauses to be added or deleted, and for variables to be inserted.

[5] Regarding machine learning and law, see Harry Surden *Machine learning and law* 89

- An outdated approach to content generation and publishing is part of the legal industry's broader vulnerability to changes in technology and new legal business models.[6] Clients are fed up with the traditional law firm. Lawyers cost too much and aren't productive enough. Lawyers haven't looked outside law to see what's happening in the world. Lawyers need to learn some new tricks, and start to catch up with everyone else.

## 2. A solution

### *2.1. Thinking about the roles that contracts perform*

Lawyers author and publish contracts as if their only purpose is to be a permanent record of the parties' bargain, to be kept in safe storage and dug out and presented to the court if absolutely necessary.

But many contracts are living documents that go through a period of evolution as they are drafted and negotiated, and then use and reuse after they are signed.

Contracts perform a number of roles, and have a variety of data and business knowledge embedded within them. The roles of contracts include:

- A platform for negotiation and collaboration. Many commercial aspects of deals are conceived or refined through the formal process of reducing the parties' deal to contractual provisions.
- A store of information about the deal. This may appear in a contract in various forms, but may include classes of information such as monetary amounts, dates, references to external events, etc.[7]
- A store of knowledge about the parties' businesses. A variety of business data may be embedded within a contract, from simple things like address and contact details, to more detailed commercial information such as insurance policies, business procedures (eg. ways of making payments), etc.
- A store of knowledge about the law. Much of a lawyer's value comes from the experience of previous transactions. The content of contracts are a mineable resource for future deals.
- A reference for the court, should a dispute arise.

---

Wash. L Rev. 87 (2014). In relation to natural language processing techniques applied to law, their accuracy levels and challenges, see Oanh Thi Tran et al., *Automated reference resolution in legal texts* 22 Artif Intell Law 29 (2014).

[6] For a discussion of the business and technological pressures affecting law firms, see Richard Susskind, *The end of lawyers? Rethinking the nature of legal services* (2010) and Michael Trotter, *Declining prospects: How extraordinary competition and compensation are changing America's major law firms* (2012).

[7] The role of contracts as stores of information and as a tool for communication is discussed in Gillian Hadfield and Iva Bozovic, *Scaffolding: Using formal contracts to build informal relations in support of innovation*, University of Southern California Law School, Law and Economics Working Paper Series Paper 144 (2012).

### *2.2. More functional contracts*

Having regard to these different roles of contracts, it is worth considering whether the current approach to authoring and publishing contracts takes full advantage of available technologies to maximize their value and usefulness.

Imagine content authored by lawyers being spun out seamlessly to clients, supervising partners and other parties in a form adapted to their needs and use. Only need to see one clause in a 200 page contract? See the clause, its history, edit it and send it around without wading through the rest.[8] Your partner can approve your clause amendments on their phone in a cafe just as easily as they can in the office. We could see genuine collaboration between teams of lawyers and their clients in legal content creation.

Have suggested definitions and clauses appear as you draft. Have the software check that you haven't defined a term then not used it, or vice versa. Hover over terms and see their meaning, change histories, relationship to other parts of the document, or other documents. Your document is effortlessly available as a precedent for later transactions.

This could be achieved within a software environment based on lawyers authoring in a document markup language like XML, HTML or LaTeX, specifically adapted to law. They should also take tools from the coder's toolbox like GitHub to help them share, edit and record change histories of documents.

This is not hard. Lawyers would need to produce content in a way unfamiliar to them now. But other professionals have been producing content in this way for years.

## 3. What this could look like

### *3.1. Key elements*

Technology elements of this solution might be:

- Pre-defined tags, which will enable the content to populate a relational database. These tags would need to be applied in a standard way by all users of the system.

[8] This is one example of how the built-in modularity of contracts as discussed in George Triantis, *Improving Contract Quality: Modularity, Technology, and Innovation in Contract Design* 18 SJLBF 177 (2013) can be exploited. Richard Susskind relates how Allen & Overy intentionally re-engineer its standard loan documents to make them for modular and more suited to the imposition of IT, Richard Susskind, *The end of lawyers: Rethinking the nature of legal services* (Oxford, 2010). Similarly, it has been noted that the functionality of online legislation resources has been substantially facilitated by the modular structure of legislation (typically containing elements such as parts, divisions, sections and clauses), able to be expressed as divisional elements able to be easily parsed and distributed amongst multiple applications, see Armin Wittfoth et al. *AustLII's Point-in-Time legislation system: A generic PiT system for presenting legislation* http://tinyurl.com/austliipit, last visited May 21 2015, and Armin Wittfoth et al. *Can one size fit all?: Austlii's point-in-time legislation project* 6 UTS L. Rev. 199 (2004). Applications such as Quip and Confluence demonstrate how collaborative authoring, tracking and commenting capabilities in a user-centred design, see http://www.quip.com and https://www.atlassian.com/software/confluence.

**Permitted disclosure of Confidential Information**

The parties must not disclose Confidential Information, unless:

(a) required by the Corporations Act section 87 or any other Law;
(b) to a Prospective Investor; or
(c) in accordance with clause 19.

Figure 1: Example clause

- An assisting interface in which lawyers can author content in a markup language (basically a text editor with additional features). The assisting interface might look like Sublime or Python IDLE. The markup language might look like XML, using the pre-defined tags.

- A web-based app which users can log into, to see a contract workflow customized to them. This would include version and changes control, allowing control of edits and tracking of the history of a document. Similar to GitHub, but with appropriate privacy and security to suit the requirements of law.

- A database management system to manage the information collected, to allow its redeployment in various forms.

The first two of these elements are discussed in sections 3.2 and 3.3 below.

*3.2. Authoring in a legal markup language*

Figure 1 is an example contract clause. Figure 2 shows how this clause might be written in a markup language.[9]

Let's consider some examples of what we could achieve by writing in a markup language like this. In this example, we have used a backslash to indicate a tag. The tags would be predefined so that they readable by a computer. I have just made up tags for illustration. Table 1 describes how each tag in this example clause could be used.

Many more options are possible, depending on user needs.[10] For example, a law firm may be responsible for producing and administering many hundreds of contracts of similar type for a single enterprise. These may have consistent

[9] There are many ways in which information can be marked up. Refer to the discussion regarding XML and other markup options in the legal context in Go Eguchi and Laurence Leff, *Rule-based XML: Rules about XML in XML to support litigation regarding contracts* 10 Artif Intell Law 283 (2002). For a detailed discussion on the possible use of XML to markup legal documents, see Lawrence Cunningham, *Language, Deals and Standards: The Future of XML Contracts* 84 Wash. L Rev. 313 (2006).

[10] Mary Abraham contemplates automated environments that are smart enough to deliver "at the moment of need the relevant precedents, practice notes, drafting templates and writing guidance, as well as pertinent information from the client file...", in Mary Abraham, *Moving beyond KM for dogs*, Legal IT Today 28 (2013).

\section[\type{Confidentiality}]{Permitted disclosure of \def{
Confidential Information}}
The parties must not disclose \def{Confidential Information}, unless:
\begin{itemize}
\item required by the \leg{Corporations Act section 87} or any other \def{
Law};
\item to a \def{Prospective Investor}; \or
\item in accordance with clause \ref{Dispute Resolution}.
\end{itemize}

Figure 2: Example clause with markup

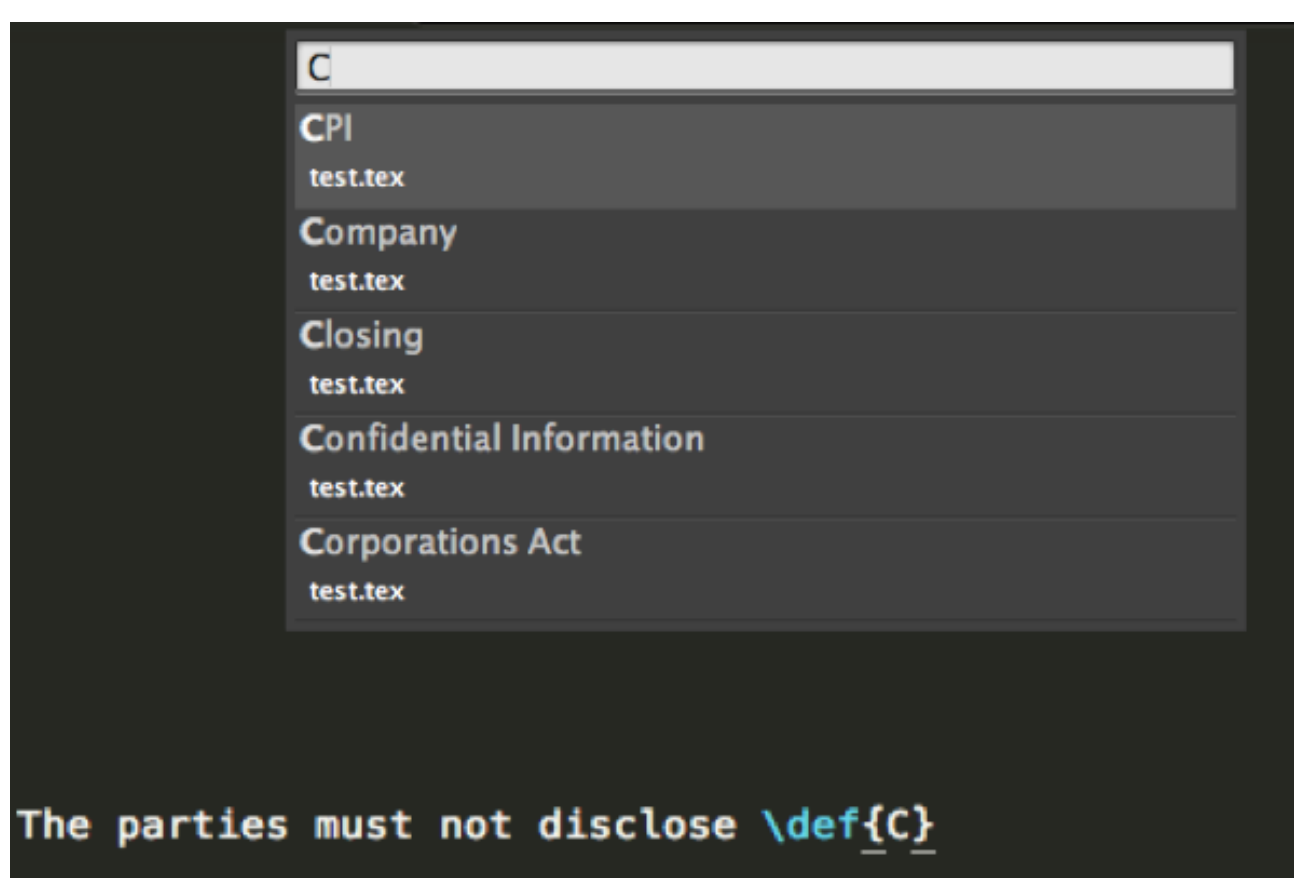

Figure 3: How a software interface can refer the author to other elements in their document, as they type

| Tag | Description |
|---|---|
| \section | Labeling sections of a contract. To allow reliable indexing of sections for uses including navigation, redeployment in other formats and later analysis. It is easy to do a hierarchy of sections to align with your style sheet, eg. \subsection, \subsubsection, etc |
| \type | Labels for content categories. You can add content to predetermined content categories (such as "Confidentiality" clauses) but not show this label in the published form |
| \def | Tagging definitions. Authoring software can check you have actually defined the term you are tagging as a definition, as shown in a screenshot of Sublime, shown in Figure 3. When you publish, you can have functionality such as being able to hover over defined terms throughout a contract and have their definition appear. |
| \leg | Tagging legislation or case names. This could have many functions, including: forcing references into a standard format (ie. "firm style"); allowing later analysis of references to legislation across documents; or more sophisticated functions like cross-checking with legislative or databases to inform the drafter of relevant information such as when the section they are referring to was last amended or how a case has been treated. |
| \itemize | Tagging lists. While this might be done purely to tell the software to format the text using a list style, it has other potential applications. For example you could automatically recall items that are listed in Confidentiality clauses in your documents (eg. find that 90% of agreements of a certain type refer to the same 10 items). |
| \or | Tagging simple operators, such as "and", "or". This could help later analysis of the context of lists. You might be also able to tag logical operators such as "if", "then" and "else", for later analysis. Operators like this could encourage lawyers to adopt more standard drafting styles. For example, the software might not compile or show errors if a document that has incorrectly structured or missing logical operators. All of this works towards legal drafting that is increasingly amenable to computational analysis and transformation. |

Table 1: Description of tags used in Figure 2

variables such as payment dates, payment calculations, jurisdiction, party details such as addresses, etc. A markup language would allow tagging of these variables to facilitate communication to relevant business units in their desired format (eg. a payment schedule). Substantial efficiencies can be gained by entering data only once.

An example of how this could be applied is to the initial contractual documentation for startups.

Funding arrangements for startups are typically explored using a waterfall or cap table in Microsoft Excel or specialised software applications,[11] where different outcomes are modeled and scenarios tested as variables are adjusted. The output from this then informs term sheet generation. Startup term sheets typically have a relatively standard and modular form[12] and a number of variables which come directly from the initial waterfall or cap table (eg. holder names, participation, valuations, discounts, dividends, liquidation preferences, etc). Having a structured term sheet template would allow variables from the waterfall or cap table to be filled automatically, and the same can be applied to later formal contract generation. It would save time and increase the accuracy of term sheet and contract generation if data could be integrated across these steps.

Structuring content in this way opens up many opportunities for manipulation that would not otherwise be possible,[13] or would be more difficult and less reliable without a consistent computer readable structure.[14] This would also free lawyers from worrying about the appearance of their legal documents. Traditional word processors are preoccupied with visual formatting, which is irrelevant to law.[15]

### *3.3. Sharing content like a coder*

A further requirement is a tool to manage versions and changes to legal documents.

---

[11] See for example the description and screenshots of waterfall and cap tables at http://tinyurl.com/waterfallandcaptables, last visited May 28, 2015.

[12] This modularity has been exploited by some law firms that offer online automatic term sheet generation, such as Cooley (`www.cooley.com`) and Wilson Sonsini (`www.wsgr.com`).

[13] Current applications of this in contract management software are discussed in George Triantis, *Improving Contract Quality: Modularity, Technology, and Innovation in Contract Design* 18 SJLBF 177 (2013). For further examples of potential applications refer to Akos Szoke et al. *Versioned linking of semantic enrichment of legal documents* 21 Artif Intell Law 485 (2013). This paper defines three levels of structure for legal documents: document standardization, conceptual modeling and logical modeling. We are proposing something similar to first level, being document standardization. Szoke describes the functionalities of this level as being: enabling semantic search, versioning, translatability, interchangeability, integrability and referencability.

[14] Harry Surden describes the difficulty for computers in processing unstructured law in Harry Surden, *Structuring US Law* (2015), http://concurringopinions.com/archives/2015/05/structuring-us-law.html, last visited May 21, 2015.

[15] Firms have strict style sheets, which can be easily applied to documents produced in a consistent markup language.

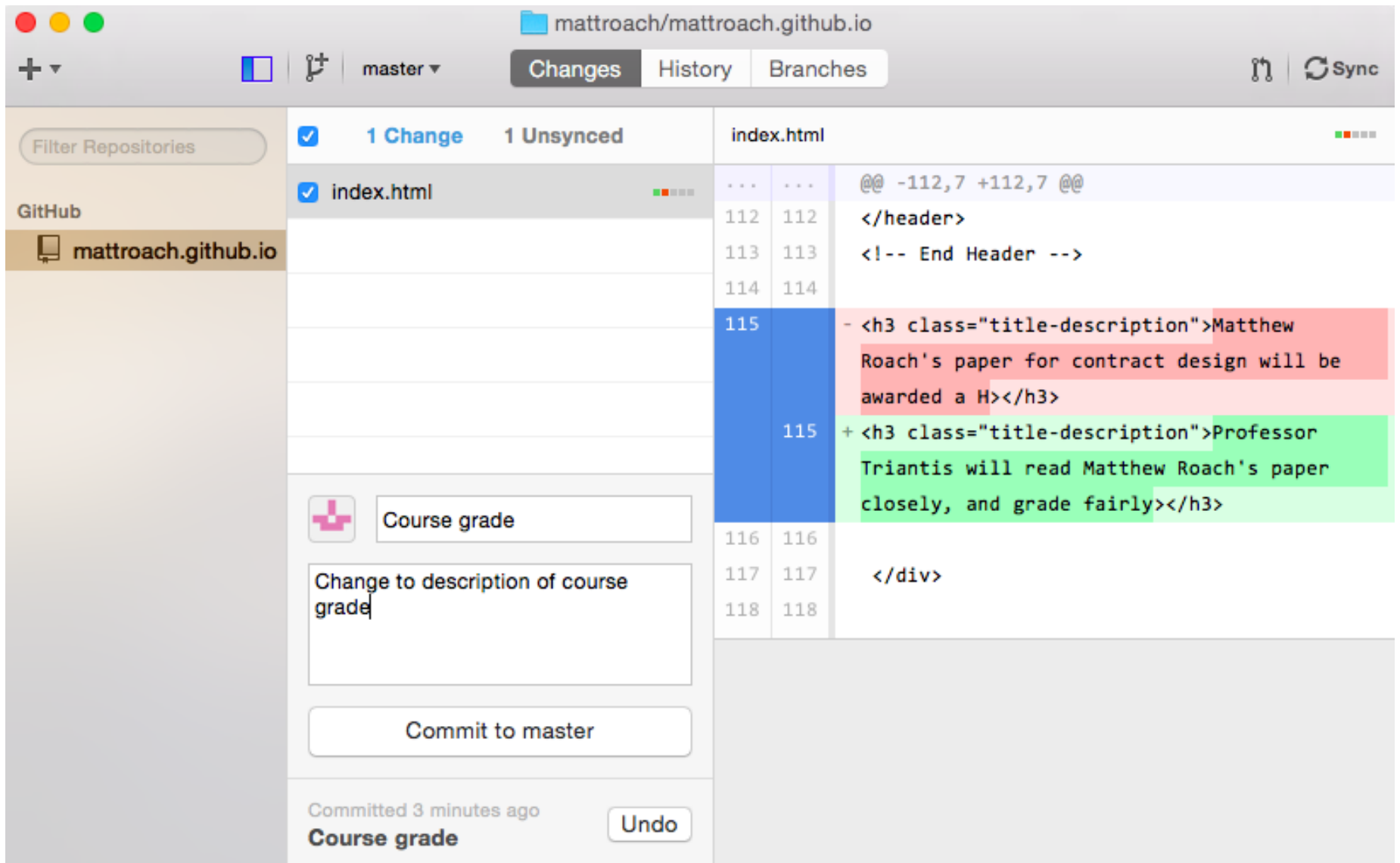

Figure 4: Example of a change to a HTML document tracked by GitHub

Coders are familiar with lengthy and complex documents produced and shared by teams, where tracking the history of changes is critical. Coders have produced better tools to do this than lawyers currently use.

Figure 4 is an example of how this looks like on the GitHub platform (red = deleted text, green = new text), with an edit made to a HTML document. Changes are identified by time and author, and can be reviewed and annotated by groups. Change history is always retained.[16]

### *3.4. What it could look like in practice: construction contracts*

This section considers how the solution might be applied to construction contracts.

Construction contracts exhibit very strong modularization, and because of the prevalence of standard forms, a high level of standardization in content.[17] Key types of clauses in a construction contract include:

- Site access.

[16] For a detailed discussion of the tasks involved in tracking changes in XML see Robin La Fontaine, Trist Mitchell and Nigel Whitaker *Representing Change Tracking in XML Markup* in XML Prague 2013 Conference Proceedings (2013), archive.xmlprague.cz/2013/files/xmlprague-2013-proceedings.pdf, last visited May 22 2015.

[17] In a survey conducted by University of Melbourne researchers in 2014, it was found that 68% of construction contracts reported upon used standard forms, John Sharkey et al. *Standard forms of contract in the Australian construction industry: Research Report*, http://tinyurl.com/unimelbreport (2014).

- Performance security.
- Design.
- Site conditions.
- Programming.
- Extensions of time.
- Payment
- Safety requirements and WHS.
- Environmental requirements.
- Completion and handover.[18]

It is possible to see from this list how authoring in a computer readable form could facilitate collaboration between lawyers and their clients through allowing identification of relevant business units for each clause or section of the contract, and dissemination and exchange of information between external lawyers, internal lawyers and business units.

Information relevant to these contractual provisions may be held by clients in different forms of data, and analyzed by them using different software programs.[19] Clients may model and runs scenarios on data which is then fed into the contract. A change to one part of the business position as set out in the contract may trigger a reassessment of other parts. Authoring contracts in a computer readable form offers the potential to allow relevant data to be maintained in a single repository to provide a source of data for different software applications, updated across all of them automatically.

An example process for a contract might look something like this:

- After recieving instructions from commercial managers, and perhaps following discussion with their external lawyers, an internal lawyer chooses the initial contract form from a precedent collection.
- The lawyer identifies modules and sections relevant to different business groups (eg. workplace health and safety, environment, insurance, accounting, etc), and sets up a web-based workflow where the relevant components of the contract are sent out to the relevant people.

[18] This list of example clauses is taken from John Sharkey et al. *Standard forms of contract in the Australian construction industry: Research Report*, http://tinyurl.com/unimelbreport(2014).

[19] For example, financial data is kept and analyzed differently to construction programming data.

- Each business unit adds in the information relevant to them, and makes such amendments to the clause as they see fit. They might also use commenting to request that the lawyers draft other amendments to achieve specific business outcomes.
- The modular format of the contract and the web-based workflow means that there can be an iterative process of reviewing and amending contract clauses before the contract is finalised.

Having an underlying computer readable structure could facilitate great flexibility in how this process could occur. It could improve productivity, and improve client satisfaction with the outcomes of the contract drafting process. Concepts that have proved successful in helping people interact in other platforms could be used here. Imagine approval of a particular clause amendment being communicated by a "like" button.

The pressing need for change in work processes in contract development and negotiation is illustrated by a University of Melbourne study on contracts. Their surveys revealed the following client views of lawyer involvement:

- Changes are driven by lawyers rather than their clients; when clients are apprised of the effects of the changes suggested by their lawyers... they often say that they do not want the changes to be made.
- Lawyers often have a poor understanding of the technical and commercial implications of their amendments, such as in advising upon contract-specific issues for insertion into the Annexure (contract particulars).[20]

The collaborative approach to contract drafting proposed here could address issues such as these by enabling input of relevant data by those most appropriately positioned to do so. It also gives the ability to comment and iterate the developing contract more rapidly.

It should also help the lawyers demonstrate their value to the client, as the lawyers will spend less time transferring and translating the business objectives into contractual form, and more time sharing insights and legal advice on the contract.

This vision of a web-based workflow contrasts sharply with the current approach to drafting in traditional word processors. For example, in a traditional word processor, the format and style encourages a mode of working where the whole contract has to go back and forth (lawyer to client, or to the other side), with very little ability to deal with modules of the contract separately.

A further potential application is the ability to directly link metrics of contract success back to clauses of the contract.[21] For example, you could have the

[20] John Sharkey et al. *Standard forms of contract in the Australian construction industry: Research Report*, http://tinyurl.com/unimelbreport(2014).

[21] This was suggested by John Moghtader, Visiting Researcher, CODEX at Stanford Law School, pers comm. May 26 2015. The value of this type of contract analysis is currently being explored by KM Standards, see Table 4 below.

authoring software report report as you draft a clause such metrics as:

- The average time taken to negotiate this clause.
- Description of changes made to the standard procedent clause in past transactions.
- The extent to which the business implementing the contract has found that this clause is not complied with or not fully complied with.[22]

## 4. Possible objections

### *4.1. Lawyers can't or won't draft like this*

A first impression may be that the tagging example given in Figure 2 is hard to read and would be annoying to draft. My response to this is:

- Not too hard to read, because the lawyer could have on their screen at the same time a simultaneous compilation of the markup text that shows it in its published form, without the tags.
- Not too hard to draft, given that lawyers are already good at writing in a highly structured way.[23] Drafting is already a slow and somewhat mechanical writing process, and this would not make it significantly more so.
- Not beyond the capacity of lawyers, given that authors in many other fields have learned to author content in markup languages. The next generation of lawyers is also increasingly familiar with software development and digital publishing, with many having done basic coding at school or university.

### *4.2. This won't be effective without industry-wide standards*

A markup language does not need to be widely adopted in order to be useful.[24] Markup languages and the software that supports them are sufficiently low-cost to develop that they can be implemented with moderate investment at the firm level,[25] although it may be more appropriate to have a software company develop the system than do it inhouse in a law firm.[26] The need to have a standard format for document exchange in a legal transaction is outdated. Online data rooms such as Ansarada[27] are a perfect example of how the party holding the pen in a transaction can determine the interface by which clients and other parties interact with transaction documents. It is submitted that it is perfectly plausible that many contracting parties could manage their contract negotiations within web-based applications that are accessible to lawyers, their clients and the other side, without needing recourse to emailing documents back and forth in traditional word processor format.

[22] For example, a contract may require something to be done in 10 business days but in practice is almost never done in this time and that is in fact broadly acceptable to the business. Lesson: why waste time negotiating for something different?

[23] In the context of US Code, Harry Surden describes law as having a strong "implicit" structure, which he contrasts with "explicit, machine-readable" structures. See Harry Surden, *Structuring US Law* (2015), http://concurringopinions.com/archives/2015/05/structuring-us-law.html, last visited May 21, 2015. I suggest that this applies also to legal contracts, albeit to a slightly lesser degree.

[24] Although there are obvious benefits if standardization were possible. In the area of drafting and publishing legislation in Australia, for example, it has been noted that the lack of consistency or uniform approach has been a major lost opportunity, Michael Rubacki, *Online legislation from Australian Governments: achievements and issues*, Paper prepared for AustLII Research Seminar, May 7 2013, http://www.austlii.edu.au/austlii/seminars/2013/1.pdf, last visited May 21, 2015.

*4.3. This can be done as well or better within traditional word processors*

Word processors commonly used by lawyers, such as Microsoft Word, have vast functionality and are highly customizable. Large law firms have sophisticated built-in style sheets, integration with precedent systems and firm databases, and firm-specific macros. Products such as ContractExpress Author provide examples of how computer readable structure can be added to and exploited by apps operating within traditional word processors.

Figure 5 is a screenshot of ContractExpress Author showing variables being added to and manipulated within a contract, and its integration with Microsoft Word. The functionality of ContractExpress Author includes:

- The ability for the user to create their own variables.
- Integration with clause libraries
- Automated insertion of repeated information (such as details about parties)
- Support integration with external databases.
- Integration with external dictionaries.
- Validation alerts such as date ranges, text lengths and text masks.[28]

[25] Lawrence Cunningham, *Language, Deals and Standards: The Future of XML Contracts* 82 Wash. L Rev. 313 (2006) argues that the very ease of creating legal-specific markup means that standards are necessary to define a single vocabulary so that legal markup languages do not become a "Tower of Babel". To see an example of a project that has implemented XML in legal contracts (in this case end user licence agreements for software), see George Bina *An XML solution for legal documents* in XML Prague 2013 Conference Proceedings (2013), archive.xmlprague.cz/2013/files/xmlprague-2013-proceedings.pdf, last visited May 22 2015.

[26] The actual cost would depend on the functionality and sophistication of the system, but nothing proposed in this paper is particuarly innovative from a software perspective. All that is unusual is applying it to law.

[27] See http://www.ansarada.com.

[28] Business Integrity, *ContractExpress Author*, `http://www.business-integrity.com/technology/contractexpress-author/`, last visited May 25 2015.

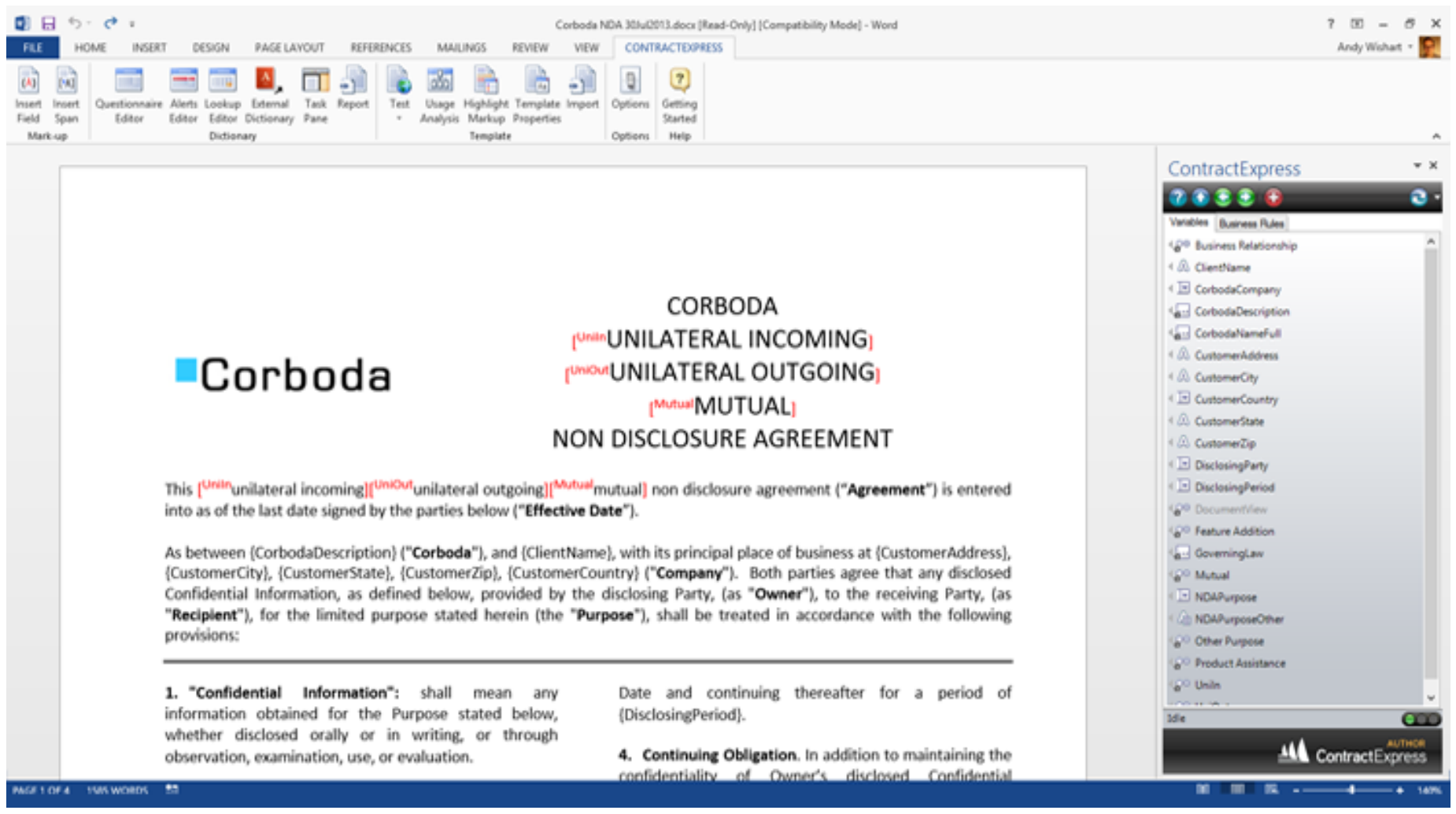

Figure 5: Screenshot of ContractExpress Author (BusinessIntegrity 2015)

Using apps to add this functionality to traditional word processors carries with it the substantial benefits of using a tool with which lawyers are already familiar and comfortable.

Can then a case be made out to depart from traditional word processors, which have the immense benefits of almost universal market saturation, are highly functional and highly customizable?

One issue with traditional word processors such as Microsoft Word is that to the extent that machine readable structure is able to be added, it is entered into a proprietary software with limited ability for reuse in other applications. The underlying format of files authored in traditional word processors is designed to remain within that software.[29]

Arguments for not using a traditional word processor include that:

- Much of the functionality of traditional word processors can be replicated in other software, and in fact the opportunity exists to intentionally choose pared down functionality that is more fit for purpose.
- If you added computer readable structure to contracts using an app integrated into a word processor, once you load it up to a repository and download it in another application you are likely to lose much of the formatting, style and track changes, as these elements of the document will be stored in the word processor's proprietary format and not accessible to

[29] The extent to which this can be overcome or to which workarounds are possible by apps such as ContractExpress Author could be further considered, but is beyond the scope of this paper.

other applications.[30]

- Writing in a more directly computer readable form using the conventions of mark-up languages seems intimidating at first, but can be learned relatively easily.

- If we are going to follow a trend of increasingly adding computer readable structure to legal content, then for efficiency reasons it seems desirable for lawyers to add the structure directly. Relying on programmers to perform these tasks adds delay and cost, and there are often enormous time pressures on contract drafting.[31]

- In my view, the user experience for writing in a markup language is better than relying on a software platform that conceals computer readable information that is relevant to the author. Lawyers have an interest in controlling the whole content of what they are authoring, and as the computer readable element of this grows in sophistication and usefulness it is increasingly desirable for lawyers to have visibility of what is occurring "behind the scenes" and in the metadata.

- A significant intangible benefit of learning to use a markup language is becoming literate with a mode of thinking and writing that is increasingly important in the modern world. My own experience with math and engineering students has been that their learning markup or programming languages leverages into other creative activities involving technology. Lawyers should be looking to develop highly deployable and relevant skills such as this to give them a better chance of adding value to their legal service and differentiating themselves from their competitors.

### *4.4. The structure is vulnerable to changes in technology*

An issue with any digital information is its vulnerability to changes in technology, which may make it obsolete or inaccessible if it is not regularly migrated to new software and new platforms.[32]

However, within a firm that has adopted particular conventions of standard markup language, the data should remain sufficiently clean and consistent to enable easy migration over time, though older files may have less functionality than newer ones. The foundation of this solution is basic text files, which it is difficult to envisage being made completely obsolete or inaccessible.

[30] Pers. comm. Tarjei Maridal, Software Developer at Adaptive Insights Australia, May 27 2015.

[31] BusinessIntegrity says that one of the benefits of their Microsoft Word app is that it reduces the need to rely on programmers to add structure to legal documents. Refer to ContractExpress product presentation.

[32] It has been argued that technological systems can guarantee no more than 50 years of access, Claire Germain, *Digitizing the World's Laws*, in Richard Danner and Jules Winterton (eds), *The IALL International Handbook of Legal Information Management* (2011) p 195.

### 5. How this compares to what others are doing

This section considers how my proposal relates to other work being done in legal document production and publishing.

Table 2 describes examples of where markup languages have been used to add structure to legal content. Table 3 describes some companies which currently use structured precedents to provide contract automation services. Table 4 describes examples of companies which undertake computational analysis of legal texts, the efforts of which could be enhanced by legal texts with existing machine readable structures.

These companies and products give interesting examples of how legal content can be manipulated and analyzed in non-traditional ways through technology. However, we were not able to find an example of the solution proposed in this paper, where practicing lawyers would use a markup language throughout the life of contract documents from initial creation to publishing.

### 6. Conclusion

Authoring legal documents in markup language offers enormous potential to enhance how legal documents are drafted, shared, and reused. While there would be a learning curve for lawyers, many other professionals have managed to do it, and lawyers have long practiced writing in a highly structured way.

Learning a new skill might even add interest to some lawyers' lives and encourage an attitude of creativity and innovation in other areas of their practice. Programming is such a useful skill that we should be optimistic about how even basic familiarity with a legal markup language can give lawyers a greater awareness of what's possible in software development, build capacity and spark the desire to make tools to meet needs that lawyers are uniquely positioned to see.

Existing systems and software used in the legal industry have adopted computational structures before. For example, many firms have structured versions of precedents to enable automated contract assembly. Contract Express and others also have structured base precedent documents. However, while these systems have structure initially, they output into normal word processing formats such as Microsoft Word usually before the contract is significantly customized and negotiated. Any structure within the precedent is lost as soon as a lawyer starts to edit it in a normal word processor. It seems to have always been assumed that lawyers could not learn to use anything else.

The field of computational law is rapidly developing, and there may be increasing demand for law to be written in formal computational structures.[33]

[33] For a general introduction to computational law refer to Michael Genesereth, *Computational Law*, http://logic.stanford.edu/classes/cs204/complaw.html, last visited May 21 2015. Also, Harry Surden, *Computable Contracts* 46 UC Davis Law Review 629 (2012) and Go Eguchi and Laurence Leff, *Rule-based XML: Rules about XML in XML to support litigation regarding contracts* 10 Artificial Intelligence and Law 283 (2002), which discusses how markup can facilitate computational reasoning and rule-based technologies, and what this might look

| | |
|---|---|
| AustLII Australasian Legal Information Institute | • One of the earliest free online legislation repositories<br>• Uses markup language to enable search and display of legislation by section, linking and indexing<br>• Has proved to be an invaluable resources for Australian lawyers because of its simplicity of access and use |
| LegalXML | • Developing open XML standards for legal documents<br>• Focused on court documents<br>• Investigating how legal arguments can be created, evaluated and compared using rule representation tools, and self-proving electronic legal information |
| Akoma Ntoso XML for parliamentary, legislative & judiciary documents | • Defines parliamentary, legislative and judiciary documents in XML formats<br>• Makes explicit the structure and semantic components of digital documents<br>• Has drafted legislative drafting guidelines that define common structural elements of legislation |

Table 2: Examples of legal XML development

| | |
|---|---|
| ContractExpress | • Integrated contract development platform<br>• Uses many of the benefits of a markup language in how it produces documents, but ultimately produces "flattened" documents in which the drafter is not involved in giving the document a meaningful structure for later computational analysis |
| legalzoom | • Offers consumer legal services through their website, including incorporation, trademark, real estate and patent services<br>• Although designed to be used by consumers, it often used by lawyers who find it to be an efficient workflow management system |
| shake SIMPLY LEGAL | • Provides free legal agreements online, from your smartphone<br>• Focuses on "tiny law" and small transactions<br>• Interface asks plain language questions before producing a contract |

Table 3: Examples of companies providing contract automation services

| | |
|---|---|
| Palantir | • Focused on making products that "make people better at their most important work"<br>• Applies a wide-range of data analytics and data platforms, along with a consulting style of working to address client problems in many industries<br>• Increasingly active in providing advice to the legal industry |
| WESTLAW® BUSINESS | • Enables automated checking of legal documents, including checking cross referencing, unpaired punctuation, defined term discrepancies, phrases and open issues |
| KMStandards | • Natural language processing and some manipulation of contract documents<br>• While this uses the benefits of a markup language, this is not visible to the user. Also, it is likely that the functionality of KM Standards could be increased if it were analyzing semi-structured documents as I am proposed, rather than the largely unstructured documents they currently use<br>• KM Standards is also a contract automation technology, which can classify contracts and create a reference standard against which to analyze other contracts, and from which users can generate new contracts |
| Lex Machina | • Predictive analytics which synthesizes case by district, outcome and by judge and gives you a percentage rating of prospects<br>• Based on computational analysis of cases, which produces multitudes of tags in case texts, none of which are applied by humans |

Table 4: Examples of companies providing legal data analytics

This is not what is proposed here, my solution is less ambitious. However, if lawyers began drafting in markup languages, there might be potential for "hybrid" contracts, which have elements of computational law and natural language provisions. Over time this could facilitate convergence between traditional legal drafting and computational law, and perhaps permit contracts that are a hybrid of computational law and natural language. This is much easier to envisage being implemented when lawyers have become familiar with computer readable languages and content creation tools that more resemble software development tools than traditional word processors.

like if marked up documents were submitted during litigation. Refer also Akos Szoke et al. *Versioned linking of semantic enrichment of legal documents* 21 Artif Intell Law 485 (2013) which contains a discussion about embedding logic structures in legal texts.